\documentclass[sigconf,natbib=true]{acmart}

\usepackage{bm}
\usepackage{enumitem}
\usepackage{multirow}

\AtBeginDocument{%
  \providecommand\BibTeX{{%
    \normalfont B\kern-0.5em{\scshape i\kern-0.25em b}\kern-0.8em\TeX}}}

\acmBooktitle{Woodstock '18: ACM Symposium on Neural Gaze Detection, June 03--05, 2018, Woodstock, NY} 
\acmPrice{15.00}
\acmISBN{978-1-4503-XXXX-X/18/06}

\usepackage{xspace}
\usepackage{balance}

\newcommand{\webtrenta}{\textsc{Web30K}\xspace}
\newcommand{\yahoo}{\textsc{Yahoo}\xspace}
\newcommand{\istella}{\textsc{Istella-S}\xspace}

\newcommand{\opponent}{Neural RankGAM\xspace}
\newcommand{\nrgam}{NRGAM\xspace}
\newcommand{\ebm}{EBM\xspace}
\newcommand{\ebminter}{EBM{$_i$}\xspace}

\newcommand{\ilmart}{{ILMART}\xspace}
\newcommand{\ilmarttwog}{{ILMART$_i$}\xspace}

\newcommand{\lmart}{{LambdaMART}\xspace}

\usepackage{acronym}
\acrodef{OURALGO}[ILMART]{Interpretable LambdaMART}
\acrodef{PIRLS}[PIRLS]{Penalized Iteratively Reweighted Least Squares}
\acrodef{ML}[ML]{Machine Learning}
\acrodef{EBM}[EBM]{Explainable Boosting Machine}
\acrodef{GAM}[GAM]{Generalized Additive Models}
\acrodef{NDCG}[nDCG]{normalized Discounted Cumulative Gain}
\acrodef{RMSE}[RMSE]{Root Mean Square Error}
\acrodef{LTR}[LtR]{Learning to Rank}
\acrodef{PR}[PR]{PageRank}
\acrodef{QUCC}[QUCC]{Query-url Click Count}
\acrodef{UCC}[UCC]{URL Click Count}
\acrodef{LIMIR}[LIMIR]{Language Model for Information Retrieval}

\begin{document}
\fancyhead{}
\title{\ilmart: Interpretable Ranking with Constrained LambdaMART}

\author{Claudio Lucchese}
\email{claudio.lucchese@unive.it}
\orcid{0000-0002-2545-0425}
\affiliation{%
  \institution{Ca' Foscari University of Venice}
  \country{Italy}
}

\author{Franco Maria Nardini}
\email{francomaria.nardini@isti.cnr.it}
\orcid{0000-0003-3183-334X}
\affiliation{%
  \institution{ISTI-CNR}
  \country{Italy}
}

\author{Salvatore Orlando}
\email{orlando@unive.it}
\orcid{0000-0002-4155-9797}
\affiliation{%
  \institution{Ca' Foscari University of Venice}
  \country{Italy}
}

\author{Raffaele Perego}
\orcid{0000-0001-7189-4724}
\email{raffaele.perego@isti.cnr.it}
\affiliation{%
  \institution{ISTI-CNR}
  \country{Italy}
}

\author{Alberto Veneri}
\email{alberto.veneri@unive.it}
\orcid{0000-0003-2094-3375}
\affiliation{
  \institution{Ca' Foscari University of Venice}
  \institution{ISTI-CNR}
  \country{Italy}
}

\renewcommand{\shortauthors}{Lucchese et al.}


\begin{abstract}
Interpretable Learning to Rank (LtR) is an emerging field within the research area of explainable AI, aiming at developing intelligible and accurate predictive models.
While most of the previous research efforts focus on creating post-hoc explanations, in this paper we investigate how to train effective and \textit{intrinsically-interpretable} ranking models.
Developing these models is particularly challenging and it also requires finding a trade-off between ranking quality and model complexity. State-of-the-art rankers, made of either large ensembles of trees or several neural layers, exploit in fact an unlimited number of feature interactions making them black boxes.
Previous approaches on intrinsically-interpretable ranking models address this issue by avoiding interactions between features thus paying a significant performance drop with respect to full-complexity models.
Conversely, \ilmart, our novel and interpretable LtR solution based on LambdaMART, is able to train effective and intelligible models by exploiting a limited and controlled number of pairwise feature interactions. Exhaustive and reproducible experiments conducted on three publicly-available LtR datasets show that \ilmart outperforms the current state-of-the-art solution for interpretable ranking of a large margin with a gain of nDCG of up to 8\%.
\end{abstract}



\maketitle


\section{Introduction}
\ac{LTR} is a vast research area focusing on learning effective models for solving ranking tasks.
Although being effective, state-of-the-art \ac{LTR} techniques do not deal with the \emph{interpretability} of the learned model, i.e., the possibility given to humans to easily understand the reasoning behind predictions made by the model.
The need for learning trustworthy models in \ac{LTR} is becoming a topic of increasing interest, which tries to address important concerns about privacy, transparency, and fairness of ranking processes behind many everyday human activities. 

For what regards the fairness of ranking algorithms, several solutions have been proposed to create fairer models, where the term \textit{fair} takes different meanings according to the metric employed~\cite{castillo_fairness_2019}.
However, focusing only on devising a fairer model does not imply to be compliant with new regulations that are emerging. An example of that is \emph{The Right of Explanation}, which allows each European citizen to claim a justification for every algorithm decision made with their data that affects their life~\cite{politou_forgetting_2018}.
Therefore, for highly-sensitive scenarios new \ac{LTR} approaches have to be investigated  to train  ranking models that do not jeopardize interpretability in the name of effectiveness.

In literature, two main lines of work dealing with the interpretation of ranking models are investigated. 
The first concerns the creation of post-hoc explanations for black-box models, while the second focuses on the design of intrinsically-interpretable models.
In the first line, the goal is the definition of techniques that allow, given a ranking model, to generate post-hoc explanations of the decisions made. 
Here, some of the explanation techniques used in regression or classification tasks have been adapted to the ranking scenario.
For example, two recent contributions~\cite{singh_exs_2019,verma_lirme_2019} present two different approaches to adapt LIME~\cite{ribeiro_why_2016} to the ranking task. 
The work by Singhs and Anand focuses only on pointwise \ac{LTR} techniques and treats the ranking problem as a classification problem, where the probability of a document to be relevant is computed with respect to its score, and a simple linear SVM is used as a local explanation model~\cite{singh_exs_2019}. 
The work by Verma and Ganguly evaluates the consistency and correctness of the explanations generated by LIME by varying the sampling strategies used to create the local neighborhood of a document~\cite{verma_lirme_2019}. 
In the same line, Fernando \emph{et al.} propose an approach to reuse SHAP~\cite{lundberg2017unified} for the ranking task~\cite{fernando_study_2019}.
Even though post-hoc explanations can be seen as a viable way to inspect and audit accurate ranking models, they can produce explanations that are not faithful to internal computations of the model, and thus they can be misleading~\cite{rudin_stop_2019}. For this reason, Zhuang \emph{et al.} pioneer the problem of developing intrinsically-interpretable ranking models~\cite{zhuang_interpretable_2021}. In detail, they propose \textit{Neural RankGAM}, an adaptation of GAM~\cite{hastie_generalized_1986} to the ranking task.
Although Neural RankGAM nicely exploits the interpretability of GAMs, the authors show that the ranking performance achieved is significantly worse than those provided by state-of-the-art black-box models.
However, it is worth noticing that in other contexts, such as in regression and classification tasks, other intrinsically interpretable models appear to be as accurate as other black-box models. A notable example of an interpretable and effective solution is \ac{EBM}, a C++ implementation of the $GA^2M$ algorithm~\cite{lou_accurate_2013}, where each component of the \ac{GAM} are learned through a mixed bagging-boosting procedure~\cite{lou_intelligible_2012}. 

\begin{figure*}
    \centering
    \includegraphics[width=\textwidth]{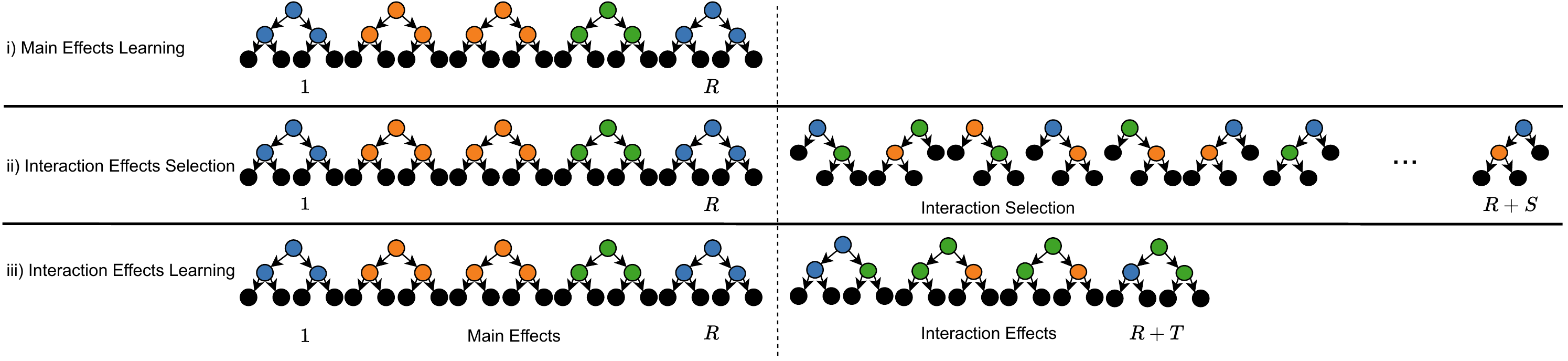}
    \caption{\ilmart learning process. Each color corresponds to a different feature used in the splits, and nodes colored in black are leaves. The trees for the main and interaction effects are represented with the same depth  for illustration purposes only.}
    \label{fig:ilmart_algo}
\end{figure*}

In this paper, we contribute to the development of intrinsically-interpretable ranking models with \ilmart, a novel \ac{LTR} solution based on LambdaMART. \ilmart creates effective and intelligible models by encompassing a limited and controlled number of pairwise feature interactions. We evaluate \ilmart on three publicly-available \ac{LTR} datasets. The results of our reproducible experiments show that \ilmart  outperforms the current state-of-the-art solution for interpretable ranking of a large margin, with a gain in terms of nDCG of up to 8\% with respect to \opponent.


\section{Interpretable \lmart}
\label{sec:ilmart}

\ilmart produces ensemble-based ranking models having the structure of a \ac{GAM}, which is considered an interpretable model~\cite{lou_intelligible_2012}. Before introducing \ilmart, we review the main aspects of \ac{GAM}s to highlight similarities and differences with respect to the newly proposed technique.

\smallskip
\noindent \textbf{\acl{GAM}}. Let $\bm{x} \in \mathbb{R}^d$ be an instance in a $d$-dimensional vector space, with $x_j$ being the value of its $j$-th feature. 
An instance $\bm{x}$ is associated with a target label $y$ generated by some unknown target function $g(\cdot)$, i.e., $g(\bm{x})=y$. In its basic form, a \ac{GAM} models the target variable $y$ as a combination of $p$ main effects and $K$ non-zero interaction effects, as defined in~\cite{hastie_generalized_1986}:
$$
\label{eq:gam}
l(\mu(\bm{x})) = \alpha + \sum_{j=1}^p s_j(x_j) + \sum_{j=1,k=1}^{j=p,k=p} s_{jk}(x_j, x_k)
$$ 
\noindent
Where $\mu(\bm{x}) = E[y|\bm{x}]$, with $y$ following a distribution of the exponential family, and $l(\cdot)$ being the so-called \emph{link function} that is able to describe the relationship between the $p$ main effects and the $K$ interactions effects with the expected value of $y$. In addition, the number of non-zero interaction effects is equal to $K$, i.e. $|\{s_{jk}(\cdot)| s_{jk}(\cdot) \neq 0 \}| = K$.
Usually, the $p$ main effects and the and the $K$ interactions are modeled through simple third order splines.
From the explainability point of view, the analyst is promptly provided with a plot of each basis function $s_j$ that clearly describes the relation between each feature and the target variable $y$, while $s_{jk}$ values represent the contribution of the interactions between feature pairs. For classification and regression tasks, (shallow) decision trees instead of splines were investigated in~\cite{lou_accurate_2013}.

\smallskip
\noindent \textbf{Interpretable \lmart}.
We now introduce \ilmart, our novel method based on \lmart to learn ranking models that are both accurate and interpretable. \ilmart presents similarities with \ac{GAM}, and builds the prediction $\hat{y}$ of a unknown target variable $y$ as follows:
\begin{equation}
\label{eq:ilmart}
\hat{y} = 
    \overbrace{
    \underbrace{\sum_{j \in {\mathcal J}} \tau_j(x_j) }_{R\ \textrm{trees}} }^{|{\mathcal J}|=p\  \textrm{main effects}}
    + 
  \overbrace{
\underbrace{\sum_{(i,j) \in {\mathcal K}} \tau_{ij}(x_i, x_j) }_{T\ \textrm{trees}} }^{ |{\mathcal K}|= K\ \textrm{interaction effects} }
\end{equation}
where $\tau_{j}(x_j)$ and $\tau_{ij}(x_i, x_j)$ are ensembles of trees. Each tree $t \in \tau_{j}(x_j)$ models a \emph{main effect} in which the only feature allowed for each split is $x_{j}$.

On the other hand, each tree  $t \in \tau_{ij}(x_i, x_j)$ models the \emph{interaction effects}, where the only features allowed in a split of tree $t$ are either $x_i$ or $x_j$.
Similarly to the notation used to describe a \ac{GAM},  $p$ denotes the number of distinct main effects $x_{i}$, whereas $K$ denotes the number of distinct interaction effects $(x_i, x_j)$.

To learn the ensemble of trees modeled by Eq. (\ref{eq:ilmart}), we adopt a modified version of the \lmart boosting algorithm. 
Unlike \ac{GAM}, we do not make any assumption on the distribution of $y$ to guide the learning process.
\noindent
The learning strategy to generate the main effects $t \in \tau_{j}(x_{j})$ and the interaction effects $t \in \tau_{ij}(x_i, x_j)$ in \autoref{eq:ilmart} is made of three steps depicted in \autoref{fig:ilmart_algo}: \emph{i)}  Main Effects Learning, which  learns a set of trees, each one working on a main effect (single feature); \emph{ii)} Interaction Effects Selection, which selects the top-$K$ most important interactions effects (feature pairs);  \emph{iii)} Interaction Effects Learning, which learns a set of trees, each one exploiting one of the interaction effects selected by the previous step.

\noindent \emph{Main Effects Learning}. 
In the initial step, illustrated on top of \autoref{fig:ilmart_algo}, we learn an ensemble  of $R$ trees modeling the main effects.
To this end, we constrain the \lmart boosting procedure to use a single feature per tree in the first $R$ boosting rounds.
In other words, at a given boosting round of \lmart, when feature $x_{j}$ is chosen for the root's split of a tree $t$ , we force the algorithm to use $x_{j}$ for all the further splits in $t$, until a stopping criterion for tree growth is reached. \lmart stops its boosting rounds when no improvements are observed on the validation set. Eventually, we obtain an ensemble of $R$ trees only using for their splits a set ${\mathcal J}$ of $p$ features, $p \leq d$ and $p \le R$,  modeling  the main effects of \autoref{eq:ilmart}.
Thus, $p$ is not an input parameter of \ilmart, but it is the number of distinct features used by the ensemble of $R$ trees modeling the main effects only.
We name this initial model \ilmart.
It is worth noting that, as a side effect of this first step, we achieve the additional effect of potentially reducing the cardinality of the features, from $d$ to $p$.

\noindent \emph{Interaction Effects Selection}.
In the second step, we still exploit the \lmart boosting procedure to \textit{select} the top-$K$ interaction effects, as illustrated in the middle of \autoref{fig:ilmart_algo}. Specifically, we continue the learning of the \ilmart model obtained so far, by enabling the interactions between all possible pairs of $p$ features identified by the previous step, according to the interaction heredity principle~\cite{cox_interaction_1984}. 
This is accomplished by constraining \lmart to fit trees with only $3$ leaves (and $2$ splits), where the $2$ splits use a pair of distinct features. 
The boosting round is stopped when the size of the set ${\mathcal K}$, composed of the distinct feature pairs used by the ensemble of $S$ trees, is exactly $K$, where $K \le \binom{p}{2}$. The $S$ trees are generated solely for the purpose of identifying the interaction effects and they are discarded afterwards.

Note that for selecting the main $K$ interaction effects, we rely on the capability of \lmart in identifying, at each boosting round, the features and the splits with the highest contributions in minimizing the loss function.

Thus, the interaction effects selected by the boosting procedure are likely be generated in order of importance. 

\noindent \emph{Interaction effects learning}.
Finally, the last step sketched on the bottom of \autoref{fig:ilmart_algo} consists in adding to the initial \ilmart model $T$ new trees learned by constraining \lmart to use only the top-$K$ feature pairs identified in the previous step. 
The number $T$ of new learned trees is not an input of algorithm, but it is chosen on the basis of the validation set. The final model, obtained by adding to \ilmart the $T$ trees working on the $K$ interaction effects is named \ilmarttwog.

\section{Experimental Evaluation}
We experimentally assess the performance of \ilmart and \ilmarttwog on three public datasets for \ac{LTR}, namely \istella, \webtrenta, and \yahoo. The \istella dataset~\cite{10.1145/2911451.2914763} includes $33$,$018$ queries with an average of $103$ documents per query. Each document-query pair is represented by $220$ features. The \webtrenta dataset~\cite{DBLP:journals/corr/QinL13} is composed of more than $30$,$000$ queries (in $5$ folds), with an average of $120$ documents per query and $136$ features per query-document pair. The \yahoo dataset~\cite{chapelle2011yahoo} is composed of two sets, including document-query pairs with $699$ features. The feature vectors of the three datasets are labeled with relevance judgments ranging from $0$ (irrelevant) to $4$ (perfectly relevant). In our experiments, we use train/validation and test splits from fold 1 of the \webtrenta dataset and data from ``set 1'' of the \yahoo dataset.

\noindent \textbf{Competitors}.
We compare the performance of \ilmart and \ilmarttwog with the following competitors:

\begin{itemize}[leftmargin=*]
    \item \opponent (\nrgam), a neural-based approach to learn GAMs for ranking. The approach exploits standalone neural networks to instantiate sub-models for each individual feature. It is the current state-of-the-art technique for learning interpretable models for ranking~\cite{zhuang_interpretable_2021}.
    \item \ebm, an efficient implementation of $GA^2M$ using a mixed bagging-boosting procedure~\cite{lou_accurate_2013}. In this case the learned model is used as a pointwise ranker, since the implementations available are only made to solve regression and classification tasks. We denote with \ebminter the \ebm model using interaction effects.
\end{itemize}

\noindent \textbf{Metrics}. We measure the performance in terms of \ac{NDCG} at three different cutoffs, i.e., $\{1, 5, 10\}$. We compute the \ac{NDCG} metric by employing exponential weighing of the relevance~\cite{burges2005learning}. By default, queries with missing relevant documents have been assigned a \ac{NDCG} $= 1.0$. The statistical significance is computed with a two-sided Fisher's randomization test~\cite{smucker_comparison_2007} with significance level $p < 0.05$. The test is computed using the RankEval library~\cite{lucchese_rankeval_sigir17}. 

\noindent \textbf{Implementation and training settings}. We implement \ilmart as an extension of the LightGBM library\footnote{\url{https://github.com/microsoft/LightGBM}}~\cite{10.5555/3294996.3295074}.
The extension works by adding the possibility to constraint the boosting procedure to use only a limited number of features per tree. 
The source code used in our experiments along with the models trained on the three public datasets is publicly available online\footnote{https://github.com/veneres/ilmart}.

The training of \ilmart and \ilmarttwog optimizes nDCG@10. We perform hyper-parameter tuning by varying the number of leaves in the trees in $\{32, 64, 128\}$ and the learning rate in $\{0.001, 0.01, 0.1\}$.
Early stopping, i.e., stopping boosting if no improvement on the validation set is observed for $100$ consecutive rounds, is used for training in step one (main effects learning) and three (interaction effects learning). The optimal \ilmart models learned with the settings above are made of $914$, $895$ and $1$,$969$ trees for the \webtrenta, \yahoo, and the \istella datasets, respectively. On the same datasets, the three optimal \ilmarttwog models are made of $1$,$365$, $1$,$215$ and $3$,$024$ trees, respectively.
\opponent is also trained by optimizing nDCG@10. We use the public implementation of NRGAM available in Keras\footnote{https://github.com/tensorflow/ranking/issues/202} and made available by the authors of the original paper~\cite{zhuang_interpretable_2021}.
We train NRGAM by using the same settings reported by the authors. In addition, we employ a batch size of $128$, and we train the networks with $3$,$000$ epochs for \webtrenta, $700$ for \yahoo and $1$,$000$ for \istella, with early stopping set to $100$ epochs. 
We also learn \ebm by optimizing nDCG@10. The performance of \ebm is fine-tuned by varying the number of outer-bags in \{5, 10, 15\} for the two versions with and without interactions. In our experiments, we employ a publicly-available implementation of \ebm\footnote{https://interpret.ml/docs/ebm.html}. For what regards the two strategies encompassing interaction effects, i.e., \ilmarttwog and \ebminter, we limit the number of maximum interactions allowed to $50$.

\begin{table}[t]
    \caption{Performance comparison in terms of nDCG. Statistically significant improvements w.r.t. \nrgam are marked with an asterisk (*). Best results are reported in bold.\label{tab:ndcg_comparison}}
    \centering
    \begin{tabular}{ccccccc}
    \toprule
    \multirow{2}{*}{Dataset} &  \multirow{2}{*}{Method} & \multicolumn{3}{c}{nDCG} & \multirow{2}{*}{$p$} & \multirow{2}{*}{$K$}\\
        & & @1 & @5 & @10 \\
    \midrule
       \multirow{5}{*}{\webtrenta} & \nrgam & 44.80 & 43.72 & 45.73 & 136 & 0\\
        & \ebm & 41.70 & 43.14 & 45.66 & 136 & 0\\
        & \ebminter & 46.51* & 45.96* & 48.01* & 136 & 50\\
        & \ilmart & 45.17 & 45.00* & 47.05* & 79 & 0\\
        & \ilmarttwog & \textbf{48.94}* & \textbf{47.76}* & \textbf{49.55}* & 79 & 46\\
        \midrule
        \multirow{5}{*}{\yahoo} & \nrgam & 67.92 & 70.13 & 75.07 & 699 & 0 \\
        & \ebm & 70.33* & 72.77* & 77.29* & 699 & 0\\
        & \ebminter & 70.22* & 72.85* & 77.36* & 699 & 50 \\
        & \ilmart & 70.57* & 72.64* & 77.21* & 153 & 0\\
        & \ilmarttwog & \textbf{70.80}* & \textbf{73.20}* & \textbf{77.57}* & 153 & 48\\
        \midrule
        \multirow{5}{*}{\istella} & \nrgam & 63.37 & 63.04 & 69.18 & 220 & 0\\
        & \ebm & 57.75 & 59.67 & 66.90 & 220 & 0\\
        & \ebminter & 62.29 & 63.05 & 69.71 & 220 & 50\\
        & \ilmart & 66.99* & 66.29* & 72.36* & 106 & 0\\
        & \ilmarttwog & \textbf{69.06}* & \textbf{68.22}* & \textbf{74.04}* & 106 & 50\\
        \bottomrule
    \end{tabular}
\end{table}

\noindent \textbf{Experimental results}.
\autoref{tab:ndcg_comparison} reports the results of the evaluation of \ilmart, \ilmarttwog and their state-of-the-art competitors.
For all the datasets considered, results show that \ilmart achieves a statistically significant improvement in terms of \ac{NDCG} for almost all the cutoffs except the case of \ac{NDCG}@1 on \webtrenta. More importantly, \ilmarttwog always achieves the best performance in terms of nDCG that always result in a statistically significant improvement over \nrgam. The best improvements of \ilmarttwog over \textit{all} the competitors is obtained  on \istella, where \ac{NDCG}@10  improves of up to 6\% w.r.t. \ebminter. On \webtrenta, we observe that \ilmarttwog improves \ac{NDCG}@10 of up to 8\% w.r.t \nrgam, although the best competitor is, in this case, \ebminter, whose performance still lags behind \ilmarttwog.
On \webtrenta, \ebminter also achieves statistically significant improvements in terms of \ac{NDCG} with respect to \nrgam. Even though \ebminter is used as a pointwise ranker, these differences confirm the efficacy of tree-based approaches in \ac{LTR} and the need to include pairwise interactions to learn more effective ranking functions.

\begin{figure}[b]
    \centering
    \includegraphics[trim={0.9cm 0.9cm 0.9cm 0.9cm},clip, width=\columnwidth]{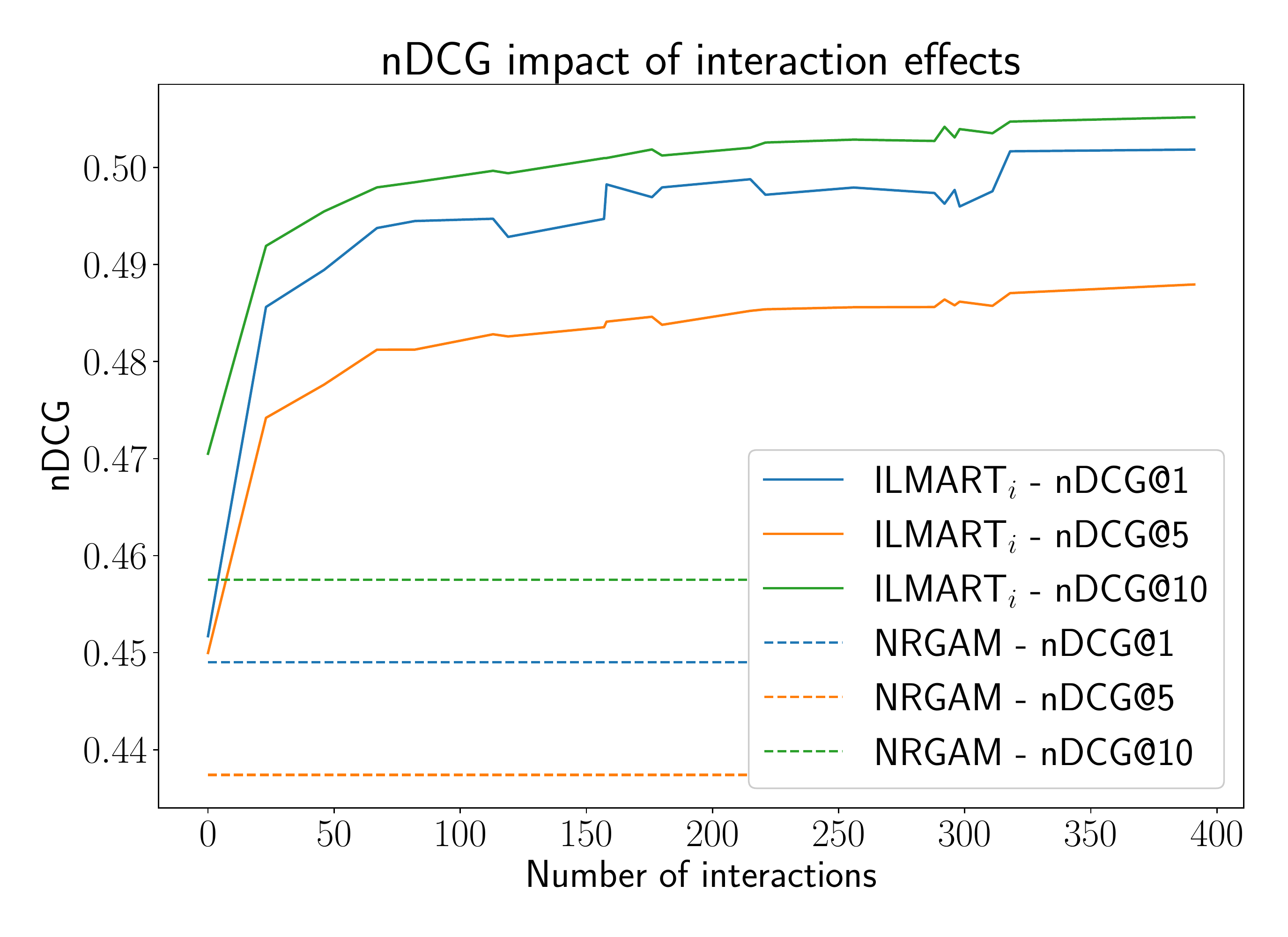}
    \caption{nDCG by varying the number of interaction effects in \ilmarttwog on the \webtrenta dataset.\label{fig:inc_inter}}
\end{figure}

\begin{figure}[t]
    \centering
    \includegraphics[trim={1.3cm 1.2 1.1 1.5},clip, width=\columnwidth]{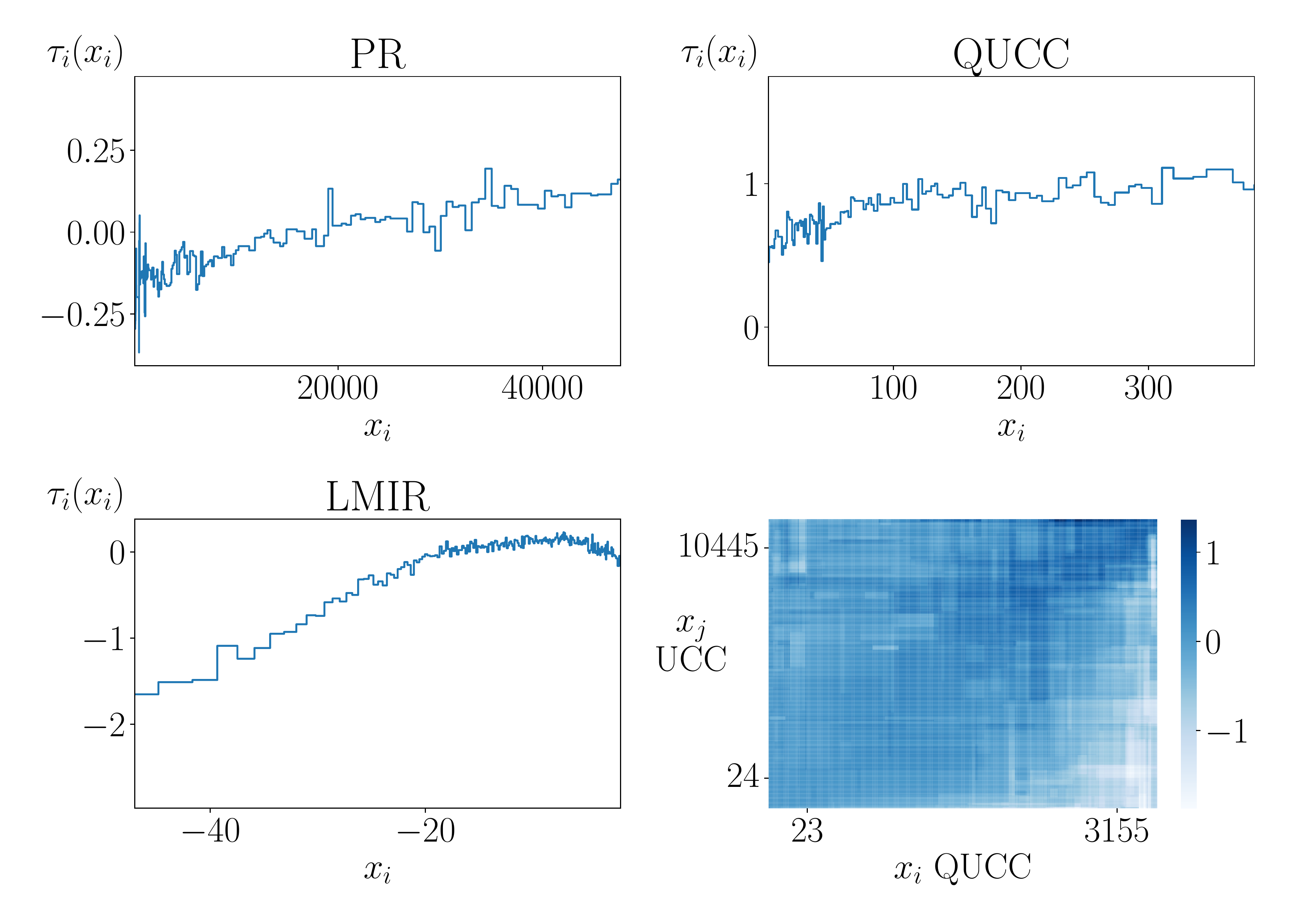}
    \caption{Plots of the four most important main and interaction effects of \ilmarttwog on the \webtrenta dataset.\label{fig:functions}}
\end{figure}

Furthermore, it is worth highlighting that our proposed techniques perform a feature selection during the \emph{main effects learning} phase, while the competitors leave the feature selection to the final user. This is an important difference between \ilmart and its competitors as a high number of features used do not allow to easily understand the global behavior of the model.
For example, the \yahoo dataset comes with a large number of features, i.e., $699$. \ilmart and \ilmarttwog effectively reduce the initial set of $699$ features to $153$ (column $p$ in Table \ref{tab:ndcg_comparison}) with, at the same time, a significant increase of the effectiveness of the learned model.
In addition, we point out that eventually \ilmarttwog may have a number of interaction effects $K$ smaller than $50$ due to the training of the new $T$ trees which may not include all the $K$ pairs selected.

We now investigate how the number of interactions added to \ilmarttwog can influence its prediction accuracy. \autoref{fig:inc_inter} compares the results of \ilmarttwog with the ones of \nrgam by reporting their results in terms of nDCG.
We observe that the ranking accuracy of \ilmarttwog increases steeply when the first interaction effects are added; then, the gain obtained by adding new interaction trees saturates. This means that \ilmarttwog is able to significantly improve the accuracy of an interpretable \ac{LTR} model by adding a small number of interaction effects, thus without hindering the overall interpretability of the model.
Even though the addition of interactions effects improves a lot the accuracy of the model, we acknowledge that the difference from the full-complexity model LambdaMART is still marked. For example, on \webtrenta an optimized black-box LambdaMART model achieves a \ac{NDCG}@10 close to 52\%.

Finally, to show how the main and interaction effects can be explored, \autoref{fig:functions} presents three main effects and one interaction effects learned with \ilmarttwog trained over \webtrenta. The three main effects selected are the ones associated with the features with highest feature importance, i.e., \ac{PR}, \ac{QUCC}, and \ac{LIMIR}, while the interaction effect between \ac{UCC} and \ac{QUCC} is the one with the highest average contribution for each value of $(x_i, x_j)$.
The 2D plots and the heatmap are built by aggregating all the trees using the same features, i.e., by representing the contribution of each $\tau_j(x_j)$ and each $\tau_{ij}(x_i, x_j)$.
From the plots, the analyst can easily understand the exact behavior of the model at varying values of each feature. For example, taking into account the interaction effect, the higher the value of \ac{UCC} and \ac{QUCC} the larger is the positive contribution of $\tau_{ij}(x_i, x_j)$.

\pagebreak
\section{Conclusions and Future Work}
We presented \ilmart, a novel and interpretable LtR solution based on LambdaMART, that allows to train effective and intelligible models. We experimentally assessed the performance of \ilmart on three public datasets. Results show that it achieves statistically significant improvements in terms of nDCG of up to 8\% w.r.t. the current state-of-the-art solution, i.e., NRGAM. We also showed that \ilmart is able to effectively exploit feature interactions without hindering the overall interpretability of the model.
With this work we laid the foundations for the development of new accurate \ac{LTR} models based on boosting methods that are also interpretable. 

As future work, we plan to investigate novel techniques based on boosting for building effective and interpretable models for ranking and new approaches for effectively identifying main and interaction effects during the learning of the ranking models, with the main goal of filling the performance gap with the black-box models.

\bibliographystyle{ACM-Reference-Format}
\balance
\bibliography{references}

\end{document}